\documentclass[12pt,reqno,a4paper,twoside]{amsart}
\usepackage{amsmath,amsthm,amstext,amscd,amssymb,amsfonts}
\newcommand{\eps}{\varepsilon}

\newcommand{\B}{\mathcal{B}}
\newcommand{\Z}{\mathbb{Z}}

\newcommand{\Bn}{\B_n(x,\eps)}

\newcommand{\Vf}{\mathcal V_f(X)}
\newcommand{\R}{\mathbb R}
\newcommand{\N}{\mathbb N}
\newcommand{\Fix}{\text{Fix}}

\renewcommand{\phi}{\varphi}

\numberwithin{equation}{section}         
\newtheorem{thm}{Theorem}[section]
\newtheorem{lem}[thm]{Lemma}
\newtheorem{defn}[thm]{Definition}
\theoremstyle{remark}

\newtheorem*{exam}{Example}

\makeatletter
\renewcommand{\@oddhead} 
{\raisebox{0pt}[\headheight][0pt]{
\vbox{\hbox to 0.99\textwidth{ \hfil \strut {\sc\small
{Fluctuation Symmetry for Chaotic Homeomorphisms
}}\hfil\thepage}\hrule}} }

\renewcommand{\@evenhead} 
{\raisebox{0pt}[\headheight][0pt]{
\vbox{\hbox to 0.99\textwidth{\sc\small \thepage\hfil \strut
Christian Maes and Evgeny Verbitskiy
\hfil 
}\hrule}}}

\makeatother
\begin{document}

\title{Large Deviations and  a Fluctuation Symmetry\\
for Chaotic Homeomorphisms}
\author{Christian Maes}

\address{ Instituut voor Theoretische Fysica, K.U.Leuven,
 Celestijnenlaan 200 D, B-3001 Leuven, Belgium}
\email{Christian.Maes@fys.kuleuven.ac.be}
\author{Evgeny Verbitskiy}
\address{ Eurandom, Technical University of Eindhoven,
P.O.Box 513, 5600 MB, Eindhoven, The Netherlands}
\email{verbitskiy@eurandom.tue.nl}

\date{\today}

\begin{abstract}
{We consider expansive homeomorphisms with the specification
property.   We give a new simple proof of a large deviation
principle for Gibbs measures corresponding to a regular potential
and we establish a general symmetry of the rate function for the
large deviations of the antisymmetric part, under time-reversal,
of the potential.  This generalizes the Gallavotti-Cohen
fluctuation theorem to a larger class of chaotic systems. }
\end{abstract}
\maketitle



\section{Introduction}

Since the beginning of statistical mechanics, there has been an
ongoing exchange of ideas between the theory of heat and the
theory of dynamical systems.  The thermodynamic formalism has
become a standard chapter for studies in dynamical systems and,
ever since Clausius, heat is understood as motion.

More recently, there has been a fruitful revival of connecting the
two theories. In particular, programs are running for
understanding the effect of nonlinearities on transport
coefficients and for defining nonequilibrium ensembles in terms of
Sinai-Ruelle-Bowen (SRB) measures, \cite{dorf,Ruelle1,GC}.  It was
also argued that the role of entropy production is, in strongly
chaotic dynamical systems, played by the phase space contraction.
Gallavotti and Cohen went on to prove a fluctuation symmetry for
the distribution of the time-averages of the phase space
contraction rate and they hypothesized that this symmetry is much
more general and relevant also for the construction of
nonequilibrium statistical mechanics.  In \cite{M} it was
emphasized that this symmetry results from the Gibbs formalism and
in the more recent \cite{MN} the general relation between entropy
production  and phase space contraction was investigated. It was
found that phase space contraction obtains its formal analogy with
the physical entropy production as source term in the potential
for time-reversal breaking.  For Anosov diffeomorphisms $f$, as
 were considered by Gallavotti-Cohen, this can be seen as
follows: the SRB distribution is a Gibbs measure for the potential
\[
\phi(x) = -\log ||Df|_{E^u(x)}||
\]
and
\[
\tilde\phi_1(x) =  -\log ||Df|_{E^s(x)}|| = -\phi(i\circ f(x))
\]
is its time-reversal. Here  $E^u(x)$ and $E^s(x)$ are the unstable
and stable subspaces of the tangent space  at the point $x$ and
$i$ is the time-reversal for which $f$ is dynamically reversible:
$i\circ f= f^{-1} \circ i$. The unstable and stable subspaces  are
not orthogonal but for Anosov systems, the angle between these
spaces is uniformly bounded away from $0$. Therefore there exists
a constant $C>0$ such that
$$
\Bigl| \sum_{k=0}^{n-1} \sigma(f^k(x)) -  \sum_{k=0}^{n-1}
(\phi+\tilde\phi_1)(f^k(x))\Bigr| \le C
$$
for all $x$ and $n\in \N$ with
\[
\sigma (x) = -\log ||Df(x)||
\]
the phase space contraction rate.  While, from \cite{M,MN}, the
most natural analogue of entropy production rate is given by the
antisymmetric part of the potential under time-reversal, that is
\[
\psi_0(x) = \phi(x) - \phi(i(x))
\]
for the purpose of computing the time-average and its large
deviations, for Anosov systems, no distinction can be made between
$\psi_0, \psi_1=\phi + \tilde\phi_1$ and $\sigma$. Within the
set-up of Gallavotti and Cohen, the ergodic averages of $\psi_0$
and of $\psi_1$ and of $\sigma$ have  exactly the same large
deviation rate function. Once this is recognized, it is natural to
generalize the fluctuation theorem of \cite{GC} to other dynamical
systems.  We believe this is interesting because it takes us away
from uniform hyperbolic behavior, which is not typical for real
physical systems. Moreover, physically relevant dynamical systems
such as billiards or systems of hard balls have singularities,
i.e., they are not everywhere differentiable.  Our proof however
of the fluctuation symmetry for expansive homeomorphisms with the
specification property relies on the corresponding thermodynamic
formalism established by Ruelle and Haydn \cite{Ru1,HayRu}. These
results are valid
 for homeomorphisms and hence do not require differentiable systems.

In sections 2 and 3 we start with some basic definitions and
results.  We do this in order to make the text, as much as
possible, self-contained.  Section 4 provides a new proof of the
large deviation principle for expansive homeomorphisms with the
specification property.   It is based on directly checking the
conditions of the G\"artner-Ellis large deviation theorem. Section
5 has the proof of the fluctuation symmetry of the time-averages
of what is denoted above by $\psi_1$ (or, equivalently, $\psi_0$).
We end with some further remarks.

\section{Expansive Homeomorphisms with Specification}

The following definitions and basic results can be found in
\cite{Bow1,HayRu,Ru1,KatHas}.

We will always assume $(X,d)$ to be a compact metric space.

\begin{defn} A homeomorphism $f:X\to X$ is called {\bf expansive}
if there exists a constant $\gamma>0$ such that if
\begin{equation}\label{expansive}
d( f^n(x), f^n(y))<\gamma\,\, \text{ for all } n\in \mathbb Z
\quad\text{ then } \quad x=y.
\end{equation}
The largest $\gamma >0$ is called the {\bf expansivity constant}
of $f$.
\end{defn}

Another important property is the following.
\begin{defn} We say that
$f:X\to X$ is a  homeomorphism with the specification property
(abbreviated to ``a homeomorphism with specification'') if for
each $\delta>0$ there exists an integer $p=p(\delta)$ such that
the following holds:
\newline
if
\begin{itemize}
\item[a)] $I_1,\ldots,I_n$ are intervals of integers, $I_j\subseteq [a,b]$
for
 some $a,b\in\Z$ and all j,
\item[b)] $\text{dist}(I_j,I_{j'})\ge p(\delta)$ for $j\not= j'$,
\end{itemize}
then for  arbitrary $x_1,\ldots,x_n\in X$ there exists a point
$x\in X$ such that for every $j$
$$d(f^k(x),f^k(x_j))<\delta\quad\text{for all }\quad k\in I_j.$$
\end{defn}
Homeomorphisms that are expansive and satisfy the specification
property, form a general class of ``chaotic'' dynamical systems.


For example, the following is an immediate corollary of Theorem
18.3.9 in \cite{KatHas}.
\begin{thm} \cite[Theorem 18.3.9]{KatHas} If $f:X\to X$ is a transitive
Anosov diffeomorphism, then $f$ is expansive and satisfies the
specification property.
\end{thm}
\section{Topological pressure, Regular Potentials and Gibbs Distributions}


\begin{defn} For every $n\in\mathbb N$ and $x,y\in X$ define a new
metric
$$ d_n(x,y)= \max_{j=0,\ldots,n-1} d(f^j(x),f^j(y)),
$$
and let $\Bn=\{y\in X:\,\, d_n(x,y)<\eps\}$ for $\eps>0$.

The set $E\subset X$ is said to be $(n,\eps)$-separated if for
every $x,y\in E$ such that $x\not=y$ we have $d_n(x,y)>\eps$.

The set $F\subset X$ is said to be $(n,\eps)$-spanning if for
every $y\in X$ there exist $x\in F$ such that $d_n(x,y)<\eps$.
\end{defn}

For a function $\phi:X\to\mathbb R$ (to be called {\bf potential})
and $x\in X$ put
$$ (S_n \phi)(x) = \sum_{k=0}^{n-1} \phi(f^k(x)).
$$

The {\bf topological pressure} is defined on the space $C(X)$ of
all continuous functions on $(X,d)$.
\begin{defn} For $n\in\mathbb N$ and $\eps>0$ define
\begin{equation}\label{statsum}
Z_n(\phi,\eps)= \sup_E \left\{ \sum_{x\in E} \exp\Bigl(
(S_n\phi)(x) \Bigr) \right\},
\end{equation}
where the supremum is taken over all $(n,\eps)$-separated sets
$E$. The pressure is then defined as
\begin{equation}\label{press_limit}
 P(\phi)=
\lim_{\eps\to 0} \limsup_{n\to \infty} \frac 1n \log
Z_n(\phi,\eps).
\end{equation}
\end{defn}

The topological entropy of $f$, denoted by $h_{top}(f)$, is by
definition the topological pressure of $\phi\equiv 0$. The
topological entropy of an expansive homeomorphism  on a compact
metric space is always finite, so is, the  topological pressure of
any continuous function. The topological pressure $P:C(X)\to\R$ is
a continuous and convex function.

 The following statement is known
as the Variational Principle \cite{Walters}.
\begin{thm} Denote by $\mathcal M(X,f)$ the set of
all $f$-invariant Borel probability measures on $X$. Let $\phi\in
C(X)$. Then $$ P(\phi) = \sup_{\mu\in\mathcal M(X,f)}
\left(h_\mu(f)+ \int \phi d\mu\right).
$$
\end{thm}
This result inspires the following definition.
\begin{defn} An element $\mu$ of $\mathcal M(X,f)$ is called an equilibrium
measure for the potential $\phi$ if
$$P(\phi) = h_\mu(f)+\int\phi\,\, d\mu.
$$
\end{defn}

The equilibrium measure for $\phi\equiv0 $ (if it exists) is
called a measure of maximal entropy.



We impose additional conditions on the class of potentials under
consideration. As we shall see later (Theorem \ref{gibbs}), the
corresponding equilibrium measures will then be {\bf Gibbs
measures}.

\begin{defn} A continuous function $\phi$ is called {\bf regular} if
for every sufficiently small $\eps>0$ there exists $K=K(\eps)>0$
such that  for all $n\in\mathbb N$
$$
 d(f^k(x),f^k(y))<\eps \text{ for } k=0,\ldots,n-1 \, \Rightarrow
\, \Bigl|(S_n\phi)(x)-(S_n\phi)(y)\Bigr|<K.
$$
The set of all regular functions 
is denoted by $\Vf$.
\end{defn}
\begin{exam}
For a hyperbolic diffeomorphism $f$, any H\"older continuous
function $\phi$ is in $\Vf$ \cite[Prop.20.2.6]{KatHas}.
\end{exam}

\begin{thm}\label{VarPrin} \cite{Bow1,Ru3,KatHas}
If $f$ is an expansive homeomorphism with specification and
$\phi\in\mathcal V_f(X)$ then there exists a unique equilibrium
measure $\mu_\phi$, i.e., $\mu_\phi$ is the unique element of
$\mathcal M(X,f)$  such that
$$ P(\phi) = h_{\mu_\phi}(f) + \int \phi d\mu_\phi.
$$
Moreover, $\mu_\phi$ is ergodic, positive on open sets and mixing.
\end{thm}

This equilibrium measure $\mu_\phi$ can be constructed from the
measures concentrated on periodic points in the following way. For
every $n\ge 1$ define a probability measure $\mu_{\phi,n}$
supported on the set of periodic points $\text{Fix}(f^n)=\{x\in
X:\,\, f^n(x)=x\}$ as follows
\begin{equation}\label{period}
\mu_{\phi,n}=\frac 1{Z(f,\phi,n)} \sum_{x\in{\text{Fix}}(f^n)}
e^{(S_n\phi)(x)} \delta_x,
\end{equation}
where $\delta_x$ is a Dirac measure at $x$ and $\displaystyle
Z(f,\phi,n)= \sum_{x\in{\text{Fix}}(f^n)} e^{(S_n\phi)(x)}$ is a
normalizing constant.

\begin{thm}\label{LimThm} \cite{Bow1,KatHas}
The equilibrium measure $\mu_\phi$ is a weak${}^*$ limit of the
sequence $\{\mu_{\phi,n}\}$, i.e., for every $h\in C(X)$
$$ \int h(x)d\mu_{\phi,n} \to \int h(x) d\mu_{\phi} \quad\text{as}\quad
n\to\infty.
$$
\end{thm}

The next result gives a ``local'' (i.e., Gibbs) description of
equilibrium measures for regular potentials, see \cite{HayRu} for
a detailed discussion.

\begin{thm}\label{gibbs}
\cite[Proposition 2.1]{HayRu},\cite[Theorem 20.3.4]{KatHas} Let
$f$ be an expansive homeomorphism with the specification property.
Let $\phi\in\Vf$ and denote its unique equilibrium measure by
$\mu_\phi$. Then, for  sufficiently small $\eps>0$, there exist
constants $A_\eps,B_\eps>0$ such that for all $x\in X$ and $n\ge
1$
\begin{equation}\label{estim}
A_\eps \le \frac {\mu_\phi\left( \{y\in
X:\,\,d(f^k(x),f^k(y))<\eps \text{ for }k=0,\ldots,n-1\}\right)}
{\exp\left( -nP(\phi)+(S_n\phi)(x)\right)} \le B_\eps.
\end{equation}
\end{thm}

We have seen that for every $\phi\in\Vf$ there exists a unique
equilibrium measure. Using (\ref{period}) and (\ref{estim}) we are
able to give necessary and sufficient conditions  on potentials
$\phi_1,\phi_2\in\Vf$ to have the same equilibrium measures
$\mu_1=\mu_{\phi_1}=\mu_{\phi_2}=\mu_2$.

\begin{thm}\label{TH_cohom}
Let $f$ be an expansive homeomorphism with specification. The
equilibrium measures $\mu_1$ and $\mu_2$ corresponding to the
potentials $\phi_1,\phi_2\in\Vf$ coincide if and only if there
exists a constant $c\in\mathbb R$ such that
\begin{equation}\label{cohom}
 (S_n\phi_1)(x) = (S_n\phi_2)(x)+nc
\end{equation}
for all $x\in\text{Fix}(f^n)$ and all $n$.
\end{thm}
\begin{proof} If (\ref{cohom}) holds for all $x\in\text{Fix}(f^n)$ and
$n$, then, by (\ref{period}), one has $\mu_{1,n}= \mu_{2,n}$ for
all $n$. Thus $\mu_1=\mu_2$.

Suppose that $\mu_1$ and $\mu_2$ coincide, and let
$\mu=\mu_1=\mu_2$. Consider ``adjusted'' potentials $\widehat
\phi_1 =\phi_1 -P(\phi_1)$ and $\widehat \phi_2=\phi_2-P(\phi_2)$.
Let $x\in\text{Fix}(f^n)$ for some $n\in\mathbb N$.  Applying
(\ref{estim}) for sufficiently small $\eps>0$ we conclude that
$$
A_\eps^{1} \exp{\bigl(\, (S_n\widehat\phi_1)(x)\,\bigr)} \le \mu(
\Bn ) \le B_\eps^{2} \exp\bigl(\, (S_n\widehat\phi_2)(x)\,\bigr).
$$
This implies that $(S_n\widehat\phi_1)(x)\le
(S_n\widehat\phi_2)(x)+C'$ for some constant $C'$ independent of
$x$ and $n$. Since $x\in\text{Fix}(f^{kn})$ for all $k\in\mathbb
N$ we have that
$$(S_n\widehat\phi_1)(x) =\lim_{k\to\infty} \frac
{(S_{kn}\widehat\phi_1)(x)}{k} \le \lim_{k\to\infty} \frac
{(S_{kn}\widehat\phi_2)(x)}{k} = (S_n\widehat\phi_2)(x).
$$
By symmetry we obtain the opposite inequality. Hence
$$ (S_n\widehat\phi_1)(x) = (S_n\widehat\phi_2)(x)
$$
for all $x\in\text{Fix}(f^n)$ and $n\in\mathbb N$. This implies
(\ref{cohom}) with $c=P(\phi_1)-P(\phi_2)$.
\end{proof}




We now recall some properties of the pressure  for expansive
homeomorphisms with specification. These facts will be used later
in the proof of the main results.

\begin{lem}\label{press_is_diff}   Suppose    $f:X\to   X$   is  an
expansive
  homeomorphism with specification. Let
  $\phi,\psi\in\Vf$.  Then the function $P(\phi+q\psi)$, $q\in\R$, is
  continuously differentiable with
  respect to $q$ and its derivative is given by
  $$
  \frac {d P(\phi+q\psi)}{dq} =\int \psi d\mu_q,
  $$
  where $\mu_q$ is the equilibrium measure corresponding to the potential
  $\phi+q\psi$.

  Moreover, $P(\phi+q\psi)$ is a strictly convex function of $q$ provided
  the equilibrium measure for the potential $\psi$ is not the measure of maximal  entropy.

  When the equilibrium measure for the potential $\psi$ is
   the measure of maximal  entropy
  one has
$$
 P(\phi+q\psi)=P(\phi)+q\int \psi d\mu_\phi
$$
for all $q\in\mathbb R$, where $\mu_\phi$ is the equilibrium
measure for $\phi$.

\end{lem}

The proof of this lemma is almost identical to the proof of Lemma
4.1 in \cite{TakVer}, and relies on the results of Walters
\cite{Walterspaper}, who showed that for expansive dynamical
systems differentiability of the pressure function $P(\cdot)$ at
$\phi$ is equivalent to the uniqueness of equilibrium measures for
$\phi$. Since the specification condition together with the
regularity condition on $\phi$ imply uniqueness of equilibrium
measures, we obtain the desired differentiability of the pressure
function.


In order to prove the Large Deviations result for expansive
homeomorphism with specification, we will have to use some results
on the
convergence
$$
\frac 1n \log Z_n(\phi,\eps)\to P(\phi)
$$
in (\ref{press_limit}).

\begin{defn} We say that $E$ is a maximal $(n,\eps)$-separated set if
it cannot be enlarged by adding new points and preserving the
separation condition.
\end{defn}
Thus every maximal $(n,\eps)$-separated set $E$ is
$(n,\eps)$-spanning as well.

The following estimates from \cite{HayRu} will be used later.

\begin{lem}\label{pressestim}
  Let $f$ be an expansive homeomorphisms and $\gamma>0$ be its expansivity
  constant. Let $\phi\in\Vf$. For every finite set $E$ put
  $$
  Z_n(\phi,E)= \sum_{x\in E} \exp\Bigl( (S_n\phi)(x)\Bigr).
  $$
\begin{enumerate}
\item If $\eps,\eps'<\gamma/2$ and $E,E'$ are the maximal $(n,\eps)$- and
  $(n,\eps')$-separated sets respectively, then one has
  $$
  Z_n(\phi,E)\le C Z_n(\phi,E'),
  $$
  where the constant $C=C(\eps,\eps')$ is independent of $n$. In
particular,
\begin{equation}\label{presslim}
  P(\phi) =\lim_{n\to\infty} \frac 1n \log Z_n(\phi,E_n),
\end{equation}
where $E_n$ are the arbitrary maximal $(n,\eps)$-separated sets.

\item If $f$ satisfies the specification property and
  $\eps<\gamma/2$ then there exists a constant $D=D(\phi,\eps)>0$ such that
\begin{equation}
  |\log Z_n(\phi,E_n) - nP(\phi)| < D
\end{equation}
for all $n$ and all maximal $(n,\eps)$-separated sets.

\item Suppose $f$ is expansive and satisfies the specification property, then
\begin{equation}\label{pressperiodic}
 P(\phi) =\lim_{n\to\infty} \frac 1n \log Z_n(\phi,\text{\rm Fix}(f^n))
\end{equation}

\end{enumerate}
\end{lem}

%
%

\section{Large deviations}

In this section we establish the Large Deviation Principle for
expansive homeomorphisms with specification and Gibbs measures
corresponding to regular potentials. In fact, one can deduce this
from more general results of Young \cite{Young} or Kifer
\cite{Kifer}. However, for our class of dynamical systems one can
easily check  the conditions of the G\"artner--Ellis theorem. For
the sake of completeness and to stand on it in the next section,
we provide the details here.

Suppose, $f:X\to X$ is an expansive homeomorphism with
specification, and $\phi,\psi$ are regular functions, i.e.,
$\phi,\psi\in\Vf$.

Let $\mu=\mu_\phi$ be an equilibrium measure for the potential
$\phi$. We will study the distribution of ergodic averages of
$\psi$ with respect to $\mu_\phi$. Namely, we will establish that
the limit
$$ \lim_{n\to\infty}
\frac 1n\log  \mu_\phi\Bigl(\Bigl\{x\in X: \,\, \frac 1n
\sum_{k=0}^{n-1} \psi(f^k(x))\in A\Bigr\}\Bigr)
$$
exists for the appropriate intervals $A\subset\R$.

Our first step will be the study of the so-called free energy
function. For every $q\in\R$ and $n\in\N$ define
$$
c_n(q) = \frac 1n \log \int \exp\Bigl( q(S_n\psi)(x)\Bigr)
d\mu_\phi.
$$
We have to prove that $c_n(q)$
 converges
for every $q$, and  that the limiting function $c(q)$ is finite
and sufficiently smooth in $q$. This is done in the next  lemma.

\begin{lem}\label{LD_lemma} For every $q\in\R$ the following limit exists
$$
     c(q) = \lim_{n\to\infty} c_n(q).
$$
The limit  $c(q)$ is also given by
\begin{equation}\label{MAIN-EQUATION}
 c(q) = P(\phi+q\psi)-P(\phi),
\end{equation}
where $P(\cdot)$ is the topological pressure.

The free energy $c(q)$ is finite,
differentiable and convex  for every q.
 It is strictly convex provided
the equilibrium measure for $\phi$ is not the measure of maximal
entropy.
\end{lem}

\begin{proof} We start by giving estimates for
$c_n(q)$, which will lead to (\ref{MAIN-EQUATION}). All other
properties of the free energy $c(q)$ will follow from the
corresponding properties of the topological pressure.

Let $\eps>0$ be sufficiently small. Let $E_n=\{x_i\}$ be any
maximal $(n,\eps)$-separated set. Since $E_n$ is a maximal
separated set, for every $x\in X$ there exists $x_i\in E_n$ such
that $x\in \B_n(x_i,\eps)$. Since $\psi$ is a regular function,
for $x\in \B_n(x_i,\eps)$ (i.e., $d_n(x,x_i)<\eps$) one has
$$    \Bigl|(S_n\psi)(x)-(S_n\psi)(x_i)\Bigr|\le K(\psi,\eps)
$$
for some constant $ K(\psi,\eps)$. Therefore,
\begin{align*}
\exp( nc_n(q))
&= \int \exp\bigl( q(S_n\psi)(x)\bigr) d\mu_\phi\\
&\le \sum_{x_i\in E_n} \int_{\B_n(x_i,\eps)}
    \exp\bigl( q(S_n\psi)(x)\bigr) d\mu_\phi\\
&\le \sum_{x_i\in E_n} \exp\bigl(
|q|K(\psi,\eps)+q(S_n\psi)(x_i)\bigr)
\mu_\phi(\B_n(x_i,\eps))\\
&\le C\exp(-nP(\phi))
\sum_{x_i\in E_n} \exp\bigl( q(S_n\psi)(x_i)+(S_n\phi)(x_i)\bigr)\\
& =  C\exp(-nP(\phi)) Z_n(\phi+q\psi,E_n),
\end{align*}
where $C=B_\eps\exp( |q|K(\psi,\eps))$, and where we have used an
upper estimate from (\ref{estim}) on the measures of balls
$\B_n(x_i,\eps)$.

To prove a similar lower estimate we use that for two different
points $x_i,x_j$ from an $(n,\eps)$-separated set $E_n$ the
intersection $\B_n(x_i,\eps/2)\cap \B_n(x_j,\eps/2)$ is empty.
Hence,
\begin{align*}
\exp( nc_n(q))
&= \int \exp\bigl( q(S_n\psi)(x)\bigr) d\mu_\phi\\
&\ge \sum_{x_i\in E_n} \int_{\B_n(x_i,\eps/2)}
 \exp\bigl( q(S_n\psi)(x)\bigr) d\mu_\phi\\
&\ge \sum_{x_i\in E_n} \exp\bigl(
-|q|K(\psi,\eps)+q(S_n\psi)(x_i)\bigr)
\mu_\phi(\B_n(x_i,\eps/2)) \\
&\ge C'\exp(-nP(\phi))
\sum_{x_i\in E_n} \exp\bigl( q(S_n\psi)(x_i)+(S_n\phi)(x_i)\bigr)\\
& =  C'\exp(-nP(\phi)) Z_n(\phi+q\psi,E_n),
\end{align*}
where $C'=A_{\eps/2}\exp( -|q|K(\psi,\eps))$, and we have used the
lower estimate from (\ref{estim}). Combining together the upper
and the lower estimates on $c_n(q)$ we obtain
$$ \frac 1n \log Z_n(\phi+q\psi,E_n) - P(\phi)+\frac {C_*}{n}
\le c_n(q)  \le \frac 1n \log Z_n(\phi+q\psi,E_n) - P(\phi)+\frac
{C^*}{n}
$$
for some constants $C_*,C^*$. Since for a sufficiently small
$\eps>0$ the limit
$$\lim_{n\to\infty} \frac 1n \log     Z_n(E_n,\phi+q\psi)
$$
exists (Lemma \ref{pressestim}) and is equal to $P(\phi+q\psi)$ we
obtain the first part of the statement.

The properties of the pressure function   $P(\psi+q\phi)$ are
given by Lemma \ref{press_is_diff}. This finishes the proof.
\end{proof}

The rate function $I(p)$ is obtained from the free energy $c(q)$
by a Legendre transform: for $p\in\R$ put
$$ I(p) = \sup_{q} ( qp - c(q)).
$$
Since $c(q)$ is differentiable, we can introduce the following
quantities
\begin{equation}\label{domain}
 \overline p = \sup_{q} {c'(q)} =\lim_{q\to +\infty} c'(q),
\quad
 \underline p = \inf_{q} c'(q) =\lim_{q\to -\infty} c'(q).
\end{equation}
Existence of the limits follows from the convexity of the free
energy $c(q)$. Standard arguments of convex analysis show that
$$ I(p) \text{ is finite for }p\in
(\underline p, \overline p)
$$
and $I(p)=+\infty$ for $p\not\in [\underline p, \overline p]$.
Moreover, since $c(q)$ is smooth and convex, $I(p)$ is also a
smooth and convex function of $p$ on $(\underline p, \overline
p)$.

Now all the conditions of the G\"artner-Ellis theorem
 \cite{Ellis-book} are
satisfied and we obtain a Large Deviations result for expansive
homeomorphisms with specification.

\begin{thm}[Large deviations]\label{LargeDev}
Let $f:X\to X$ be an expansive homeomorphism with specification.
Let $\phi,\psi\in\Vf$, and let $\mu_\phi$ be the Gibbs measure for
$\phi$. Assume that $\mu_\psi$ is not the measure of maximal
entropy. Then  there exists a smooth real convex function $I$ on
the open interval $(\underline p,\overline p)$ such that, for
every interval $J$ with $J\cap (\underline p,\overline p)\ne
\emptyset$,
$$ \lim_{n\to\infty} \frac 1n \log \mu_\phi\Bigl(\Bigl\{x:\, \frac
1n(S_n\psi)\in J\Bigr\}\Bigl) = -\inf_{ p\in J\cap
  (\underline p,\overline p)} I( p).
$$
\end{thm}

\section{Fluctuation Symmetry}

 Choose some regular potential
$\phi\in\Vf$ and let $\mu_\phi$ be the corresponding Gibbs
measure. As always, $f:X\to X$ is an expansive homeomorphism with
specification on a compact metric space $(X,d)$.

We make two assumptions:

\begin{itemize}

\item[{\bf A)}] {\bf Reversibility.} There exist a homeomorphism $i:X\to X$, preserving
the metric $d$  such that
$$
i\circ f \circ i= f^{-1}\quad\text{and}\quad i^2=identity.
$$
\end{itemize}

Fix an integer $k$ and define $\tilde\phi_k$ and $\psi_k$ as
follows
$$ \tilde\phi_k(x) = -\phi(i\circ f^k(x)),\quad \psi_k(x) = \phi(x)+\tilde\phi_k(x).
$$
From the point of statistical mechanics, the most natural choice
is for $k=0$.  Yet, to connect with phase space contraction for
Anosov diffeomorphisms it is natural to take $k=1$. While the
proofs remain valid and unchanged for all choices of $k$, we
choose to present the rest for $k=1$ and we simply write
$\tilde\phi =\tilde\phi_1, \psi =\psi_1$. Note that $\tilde\phi$
and $\psi$ are also regular potentials.

\begin{itemize}
\item[{\bf B)}] {\bf  Dissipativity.} Assume that
the equilibrium measure for $\psi\in\Vf$  is not the measure of
maximal entropy.
\end{itemize}

The assumption B can be viewed as a generalization of the
corresponding dissipation condition of \cite{GC} or
\cite{Ruelle1}: If the equilibrium measure for
$\psi=\phi+\tilde\phi$ is not the measure of maximal  entropy,
then
$$
 \int \psi\ d\mu_\phi>0
$$
which expresses the breaking of time-reversal symmetry.
In fact these conditions are equivalent as we will show in Theorem
\ref{Dissip} below.
 Assumption B is quite natural because only
 under this assumption can one
talk about  a non-trivial fluctuation symmetry.

To understand the role of assumption A, it is instructive to make
the following calculation.  Recall the definition of the
approximants $\mu_{\phi,n}$ of (\ref{period}), see Theorem
\ref{LimThm}.  Denote by
\[
\mathbb{E}_n(g) = \frac 1{Z(f,\phi,n)}
\sum_{x\in{\text{Fix}}(f^n)} e^{(S_n\phi)(x)} g(x)
\]
the expectation of a function $g$ with respect to $\mu_{\phi,n}$.
Let $g(x) = G(x,f(x),\ldots,f^{n-1}(x))$ for some $G$ on $X^n$ and
define $g^\star(x) = G(i\circ f^{n-1}(x),\ldots,i\circ f(x),i(x))
= g(i\circ f^{n-1}(x)) $.
Then, by a change of variables that leaves the set Fix$(f^n)$
globally invariant (under assumption A),
\[
\mathbb{E}_n(g^\star) = \frac 1{Z(f,\phi,n)}
\sum_{x\in{\text{Fix}}(f^n)} e^{(S_n\phi)(f^{1-n}\circ i(x))} g(x)
\]
Again by reversibility and by the definition of $\tilde\phi$,
 \[
\sum_{k=0}^{n-1} \phi(f^{k+1-n}\circ i(x)) = - \sum_{k=0}^{n-1}
\tilde\phi(f^{n-k-2}(x))
\]
so that for $x\in$Fix$(f^n)$, $S_n\phi(f^{1-n}\circ i(x)) -
S_n\phi(x) = -S_n\psi(x)$.  We conclude that
\[
\mathbb{E}_n(g^\star) = \mathbb{E}_n(g e^{-S_n\psi})
\]
or,  $S_n\psi$ can be viewed as the logarithmic ratio of the
probability of a trajectory and the probability of the
corresponding time-reversed trajectory, see \cite{M,MN}.  These
basic identities drive the following

\begin{thm}[Fluctuation Symmetry]\label{FLUCT-TH}
Assume A)-B). There exists $p^*>0$ such that, if $|p|<p^*$, then
$$      \lim_{\delta\to 0} \lim_{n\to\infty}\frac 1{n}\log \frac
        {
        \mu_\phi(\{x: \sigma_n(x)\in(p-\delta,p+\delta)\})
        }
        {
        \mu_\phi(\{x: \sigma_n(x)\in(-p-\delta,-p+\delta)\})
        }
        =p,
$$
where
$$ \sigma_n(x) = \frac 1{n}\sum_{k=0}^{n-1} \psi\bigl(f^k(x)\bigr).
$$

\end{thm}

\begin{proof} The fluctuation symmetry can be expressed in terms
of the rate function $I$ of Theorem \ref{LargeDev}; we must show
it has the symmetry
\begin{equation}\label{fluct}
   I(-p)-I(p)=p \quad\text{for}\quad |p|<p^*.
\end{equation}
Since $I(p)$ is the Legendre transform of
$c(q)=P(\phi+q\psi)-P(\phi)$, the symmetry (\ref{fluct}) must be
reflected in a certain symmetry of the free energy $c(q)$. We
claim that
\begin{equation}\label{press}
   c(q)=c(-1-q)\quad\text{for all}\quad q\in\R.
\end{equation}
Assume for a moment that (\ref{press}) is true. Then from
(\ref{domain}) we conclude that
$$
\underline p   = -\overline p,
$$
Since under the assumptions of the theorem $c(q)$ is not
identically equal to a constant, we obtain $\underline p \ne
\overline p$. Hence, the domain of the rate function $I(p)$ is a
symmetric interval containing zero. Let $p^* = \overline p = -
\underline p$. For every $p$ in $(-p^*,p^*)$,  $I(p)$ is finite
and satisfies (\ref{fluct}):
\begin{align*}
I(-p)&=\sup_{q\in\R}\Bigl( (-p)q-c(q)\Bigr)=
\sup_{q\in\R}\Bigl( p(-q)-c(q)\Bigr)\\
&=\sup_{q\in\R}\Bigl( pq-c(-q)\Bigr) =\sup_{q\in\R}\Bigl(
pq-c(-1+q)\Bigr)\quad\quad\quad\quad
(\text{by }(\ref{press}))\\
&=\sup_{q\in\R}\Bigl( p(-1+q)-c(-1+q)\Bigr)+p =
\sup_{q\in\R}\Bigl( pq-c(q)\Bigr)+p \\
&= I(p)+p.
\end{align*}
Now let us prove (\ref{press}), or, what amounts to the same thing
(see Lemma \ref{LD_lemma}),
\begin{equation}\label{Cond2}
  P\bigl( (q+1)\phi +q\tilde\phi\bigr) = P\bigl(
  -q\phi-(1+q)\tilde\phi\bigr),\quad\forall
q\in\R,
\end{equation}
where the topological pressure $P(\cdot)$ is obtained from
(\ref{pressperiodic}).

The `time reversing' homeomorphism $i$ maps the set $\Fix(f^n)$ to
itself. For $x\in \Fix(f^n)$ one has
\begin{gather}
\sum_{k=0}^{n-1} (1+q)\phi(f^k(x))+q\tilde\phi(f^k(x))=
\sum_{k=0}^{n-1} -(1+q)\tilde\phi(i\circ f^{k+1}(x))-q\phi(i\circ f^{k+1}(x))\notag\\
= \sum_{k=0}^{n-1} -(1+q)\tilde\phi(f^n\circ i\circ f^{k+1}(x))-
q\phi(f^n\circ i\circ f^{k+1}(x))\notag\\
= \sum_{k=0}^{n-1} -(1+q)\tilde\phi(f^{n-k-1}\circ i (x))-
q\phi(f^{n-k-1}\circ i (x))\notag\\
=\sum_{m=0}^{n-1} -(1+q)\tilde\phi( f^m\circ i(x)) -q\phi(f^m\circ
i(x)).\notag
\end{gather}
Now, taking into account that $i$ is a bijection on $\Fix(f^n)$,
we obtain that
$$
\sum_{x\in \Fix(f^n)} \exp\bigl( \sum_{k=0}^{n-1} \phi(f^k(x))
+q\psi(f^k(x))\bigr) = \sum_{x\in \Fix(f^n)} \exp\bigl(
\sum_{k=0}^{n-1} \phi(f^k(x)) - (1+q) \psi(f^k(x))\bigr),
$$
 implying $  c(q)=c(-1-q)$ and  finishing the proof.
\end{proof}

\begin{thm}[Dissipativity conditions]\label{Dissip}
Let $f$, $i$, $\phi$, and $\tilde\phi$  be as above. Then the
following conditions are equivalent:
\begin{itemize}
\item[1)] the equilibrium measure for $\psi=\phi+\tilde\phi$ is not
 the measure of maximal entropy;
\item[2)] the equilibrium measure $\mu_\phi$ for $\phi$
 is not  the equilibrium measure for $-\tilde\phi$;
\item[3)]
$\displaystyle  \int \psi\ d\mu_\phi >0. $
\end{itemize}
\end{thm}

\begin{proof}
We first show the equivalence of conditions 1) and 2). According
to Theorem \ref{TH_cohom}, the equilibrium measure for $\psi$ is
the measure of maximal entropy if and only if there exists a
constant $c_1$ such that
\begin{equation}\label{cond1}
  \sum_{k=0}^{n-1} \psi( f^k(x)) = n c_1
\end{equation}
for all $n\in \N$ and every $x\in \Fix(f^n)$. Similarly, $\phi$
and $-\tilde\phi$ have the same equilibrium measure if and only if
for some constant $c_2$ one has
 \begin{equation}\label{cond2}
  \sum_{k=0}^{n-1} \phi( f^k(x)) =-\sum_{k=0}^{n-1} \tilde\phi( f^k(x))+ n c_2,
\end{equation}
again for all $n\in \N$ and every $x\in \Fix(f^n)$. Clearly, since
$\psi=\phi+\tilde\phi$, (\ref{cond1}) and (\ref{cond2}) are
equivalent.

To show that the second and the third condition are equivalent,
first recall that in the proof of Theorem \ref{FLUCT-TH} we have
established (\ref{Cond2}), and in particular, the following
equality
\begin{equation}\label{equat_pr1}
  P( \phi) = P(-\tilde\phi).
\end{equation}
Now, since $\mu_\phi$ is an equilibrium measure for $\phi$, one
has
$$
 P(\phi) = h_{\mu_\phi}(f)+\int\phi d\mu_\phi.
$$
On the other hand, applying the Variational Principle to
$-\tilde\phi$, we conclude that
$$
  P( -\tilde\phi) \ge h_{\mu_\phi}(f) -\int \tilde\phi d\mu_\phi,
$$
with equality if and only if $\mu_\phi$ is the equilibrium measure
for $-\tilde\phi$.

Therefore
$$
\int \psi\ d\mu_\phi \ge P(\phi)- P(-\tilde\phi)=0.
$$
with equality if and only if $\mu_\phi$ is the equilibrium measure
for $-\tilde\phi$. This finishes the proof.
\end{proof}
\section{Concluding remarks.}

\noindent {\bf 1)}  As mentioned already in the introduction, the
Fluctuation Theorem of Gallavotti and Cohen says that the large
deviation rate function $I(p)$ for the time-averages of the phase
space contraction in the SRB measure for dissipative and
reversible Anosov diffeomorphism has the symmetry:
$$
    I(-p) - I(p) = p.
$$
Our results are valid under a greater generality: not only for SRB
measures, but also for the Gibbs measures for expansive
homeomorphisms with the specification property. Phase space
contraction then gets replaced by the antisymmetric part of the
potential under time-reversal which is essentially the same as the
phase space contraction rate for Anosov systems. The original
proof of Gallavotti and Cohen used Markov partitions (symbolic
dynamics). Clearly, this is not an option for us. On the other
hand, Ruelle in \cite{Ruelle1} gave a proof, again for Anosov
systems, based on shadowing and one can check that the argument
from \cite{Ruelle1} goes through without many substantial
modifications in the case of expansive homeomorphisms with
specification. Our approach is still different and even simpler:
it is different, physically, because we concentrate on the
antisymmetric part of the potential under time-reversal and
mathematically, we obtain a symmetry of the rate function $I(p)$
directly from the properties of the topological pressure.   It is
important that in this generalization, we can also treat
continuous (therefore, not necessarily differentiable)
transformations.

\medskip
\noindent {\bf 2)} Transitive Anosov systems are expansive and do
satisfy the specification property.  One can find examples of
smooth expansive dynamical systems with the specification
property, which are not Anosov, see e.g. \cite{AOKI}.
Unfortunately, we were not able to find any interesting reversible
examples. However, this is quite a typical situation in the field:
there are also not very many examples of reversible dissipative
Anosov systems. Nevertheless, we think that the validity of the
fluctuation symmetry for a larger class of dynamical systems is a
step forward. The main reason was already mentioned: uniform
hyperbolic behavior and everywhere differentiability is not
typical for real physical systems. Secondly, the definition  of
regular potentials can be extended to include discontinuous
functions which satisfy the key property
$$
 d(f^k(x),f^k(y))<\eps \text{ for } k=0,\ldots,n-1 \, \Rightarrow
\, \Bigl|\sum_{k=0}^{n-1}\phi(f^kx)-\sum_{k=0}^{n-1}
\phi(f^ky)\Bigr|<K.
$$
It is important to understand whether hyperbolic systems with
singularities, such as hard ball systems or billiards, satisfy
this condition for natural potentials. If this is indeed the case,
then one immediately obtains the fluctuation symmetry for such
systems as well.

\section*{Acknowledgment}
We are grateful to David Ruelle and Floris Takens for helpful
discussions. E.V. was partially supported by NWO grant  613-06-551
and EET grant K99124.

\end{document}